\documentclass[structabstract]{aa}
\usepackage{graphicx}
\usepackage{txfonts}

\newcommand{\mincir}{\raise
  -2.truept\hbox{\rlap{\hbox{$\sim$}}\raise5.truept \hbox{$<$}\ }}
\newcommand{\magcir}{\raise
  -2.truept\hbox{\rlap{\hbox{$\sim$}}\raise5.truept \hbox{$>$}\ }}
\newcommand{\siml}{\raise
  -2.truept\hbox{\rlap{\hbox{$\sim$}}\raise5.truept \hbox{$<$}\ }}
\newcommand{\simg}{\raise
  -2.truept\hbox{\rlap{\hbox{$\sim$}}\raise5.truept \hbox{$>$}\ }}

\begin{document}
\title{Tomographic weak-lensing
shear spectra from large N-body and hydrodynamical simulations}


\author{Luciano Casarini \inst{1}, Silvio A. Bonometto \inst{~2,3,4},
  Stefano Borgani \inst{~2,3,4}, Klaus Dolag \inst{~5,6},Giuseppe
  Murante \inst{~2,7},\\ Marino Mezzetti \inst{~2,4}, Luca Tornatore
  \inst{~2}, Giuseppe La Vacca \inst{~8,9} }

\institute{$^1$ -- Departamento de F\`\i sica, UFES, Avenida Fernando Ferrari
  514, Vit\`oria, Esp\`\i rito Santo,  Brazil \\ $^2$ -- Department of
  Physics, Astronomy Unit, Trieste University, Via Tiepolo 11, I 34143
  Trieste, Italy \\ $^3$ -- I.N.F.N. -- Sezione di Trieste, Via
  Valerio, 2 I - 34127 Trieste, Italy \\ $^4$ -- I.N.A.F. --
  Astronomical Observatory of Trieste, Via Tiepolo 11, I 34143
  Trieste, Italy \\ $^5$ -- Universit\"atssternwarte M\"unchen,
  M\"unchen, Germany\\ $^6$ -- Max-Planck-Institut f\"ur Astrophysik,
  Garching, Germany\\ $^7$ -- I.N.A.F. -- Astronomical Observatory of
  Torino, Via Tiepolo 11, I 34143 Trieste, Italy \\ $^8$ -- Physics
  Dep.~G.~Occhialini, Milano--Bicocca University, Piazza della Scienza
  3, I 20126 Milano, Italy \\ $^9$ -- I.N.F.N. -- Sezione di
  Milano--Bicocca, Piazza della Scienza 3, I 20126 Milano, Italy \\ }


\abstract{Forthcoming experiments will enable us to determine
  tomographic shear spectra at a high precision level. Most
  predictions about them have until now been biased on algorithms yielding
  the expected linear and non--linear spectrum of density
  fluctuations. Even when simulations have been used, so-called Halofit
  (Smith et al 2003) predictions on fairly large scales have been needed.} 
  {We wish to go beyond this limitation.} {We perform N--body and hydrodynamical
  simulations within a sufficiently large cosmological volume to allow a
  direct connection between simulations and linear spectra. While
  covering large length-scales, the simulation resolution is good enough to 
  allow us to explore the high--$\ell$ harmonics of the cosmic shear (up to 
  $\ell \sim 50000$), well into the domain where baryon physics becomes
  important. We then compare shear spectra in the absence and in presence of various
  kinds of baryon physics, such as radiative cooling, star formation,
  and supernova feedback in the form of galactic winds.}{We distinguish 
  several typical properties of matter fluctuation spectra in
  the different simulations and test their impact on shear spectra.
  }{We compare our outputs with those obtainable using approximate
  expressions for non--linear spectra, and identify substantial discrepancies 
  even between our results and those of purely N-body results.
  Our simulations and the treatment of their outputs however enable
  us, for the first time, to obtain shear results taht are fully independent
  of any approximate expression, also in the high--$\ell$
  range, where we need to incorporate a 
  non--linear power spectrum of density perturbations, and
  the effects of baryon physics. This will allow us to
  fully exploit the cosmological information contained in future
  high--sensitivity cosmic shear surveys, exploring the physics of cosmic 
  shears via weak lensing measurements.}



\authorrunning{Casarini {\it et al.}}
\titlerunning{Shear spectra from large simulations}
\maketitle

\section{Introduction}
Dark energy (DE) is the most remarkable finding but largest outstanding uncertainty in cosmology today. Cosmic microwave background (CMB) spectra firmly constrain
its contribution to the cosmic mass density budget, but its
equation of state, $w(z)$, can be constrained only by 
measures of the mass density field $\rho({\bf x},z)$, at low $z$. It is 
therefore important that
the analysis of tomographic weak lensing spectra, which are directly sensitive
to the whole matter distribution, can be suitably translated into
information on the density fluctuation spectrum
\begin{equation}
\label{pk}
P(k,z) = \langle | \delta(k,z) |^2 \rangle
\end{equation}
and its redshift dependence. Here $\delta(k,z)$ is the Fourier
transform of the matter fluctuation field 
$\epsilon({\bf x},z) = 
\rho({\bf x},z)/\bar \rho - 1$. 

This paper is therefore devoted to examining the relation between
$P(k,z)$ and the tomographic weak lensing spectra $P_{ij}(\ell)$,
which are defined below. This possibility has been debated by various
authors and discussed by the Dark Energy task force (DEFT: Albrecht et
al. 2006), while cosmic shear measures have already been performed by
using either large area surveys carried out with ground--based
telescopes (e.g., Hoekstra et al. 2006, Fu et al. 2008) or narrower
area surveys with high quality imaging from HST (e.g., Massey et
al. 2007, Schrabback et al. 2010). Observational campaigns to
ultimately perform the systematic mapping of tomographic cosmic shear
have been proposed for future missions, such as the recently approved
ESA Euclid satellite (Laureijs et al. 2011).

An essential point is that, according to Huterer \& Takada (2005),
 spectral predictions suitable to exploit lensing data need a
  precision $\cal O$$(1\, \%)$ in $P(k,z)$. In turn this means that
  cosmological parameter variations causing a shift $\cal O$$(1\, \%)$
  in $P(k,z)$ could become appreciable. More recently, Hearin et
  al. (2011) reconsidered the whole question, showing that the
  precision requirements depend on $\ell_{max}$ (the highest multipole
  included in the lensing survey), the precision achieved in
  photometric redshift (photo--z, hereafter) measurements, and the
  simultaneous use of other information,  such as degree scale clustering
  data.  

The relation between $P(k,z)$ and $P_{ij}(\ell)$ can be studied by
designing simulations, and it is now clear that any attempt to
  measure the DE state equation $w(a)$ requires spectral predictions
  even more precise than $1\, \%$. Accordingly, we develop
simulations here to enable us (i) to connect their spectra on large
scales with linear predictions, and (ii) to provide predictions
  up to $\ell \simeq 20000-30000$, a range that experiments will test
but no simulation ever yet explored  without simultaneously using
  spectral approximations, which unavoidably affect the normalization
  of spectral predictions, even at large $\ell$.  Our objective (i)
is achieved by using a box of $410\, h^{-1}$Mpc on a side. Objective
(ii) requires a sufficient large dynamical range, appropiate
techniques to calculate the high--$k$ spectra, and an accurate
evaluation of the numerical noise, which is initially present on the
scale of the grid where the initial conditions are defined.

The cosmology that we use to test this set of techniques is a Lambda cold
dark matter ($\Lambda$CDM)
model, for which the DE state equation is $w(z) \equiv -1~$.  
We recall that, by using a set of $\Lambda$CDM N--body simulations in 
boxes of lengths of $\sim 240\, h^{-1}$Mpc, consisting of 
256$^3$ particles, Smith et al. (2002, see also Jenkins et al. 1998)
produced numerical expressions for $P(k,z)$, based on the halo model,
claiming a precision of $\cal O$$(\pm 3\, \%)$.  These expressions, dubbed
{\it Halofit} (hereafter HF), have also been inserted into linear
calculations and, although imprecise even for $\Lambda$CDM (Casarini
et al. 2009, Hilbert et al. 2009, Heitmann et al. 2010), have been
tentatively used to work out shear spectra even for cosmologies with
DE state equations $w \neq -1$.  As might be expected, this
  extension outside the assigned range of validity causes misleading
predictions, namely when parameter estimates, performed using
tomographic shear spectra, are correlated (e.g. Casarini et al. 2011a,
Seo et al. 2011).

Besides these problems, going beyond {\it HF} expressions is a must
when exploring scales where baryon physics becomes important. In this
paper values of $k > 2$--$3\, h$Mpc$^{-1} $ are extensively treated,
as they correspond to a spectral range where cosmic shear will be
accurately measured by future surveys. Accordingly, the simulations
used in this work treat baryon physics by including adiabatic cooling,
star formation, and energy feedback from supernovae (SNe). Comparing
them with simulations run with purely gravitational forces, or in the
presence of simplified baryon physics, will enable us to follow the
impact of different physical effects on both fluctuation and cosmic
shear spectra and gauge the level of approximation in shear spectra
when using {\it HF}.  Several authors have dealt with these problems
in the past few years.  Jing et al. (2006) were the first to estimate
the impact of radiative gas cooling on fluctuation spectra measured
from hydrodynamical simulations, finding shifts of $\cal O$$(10\, \%)$
in the 1--10$\, h$Mpc$^{-1}$ $k$--range. Rudd et al. (2008), using a
box of 60$\, h^{-1}$Mpc and $(2 \times)256^3$ particles, compared
spectra obtained using pure gravitational dynamics, with those based
on either non--radiative or radiative baryon physics.  Casarini et
al. (2011a, 2011b) made use of boxes of 256 and 64$\, h^{-1}$Mpc on a
side, with $(2 \times) 256^3$ particles, to analyse matter fluctuation
spectra and shear spectra in models with different DE state equations,
and studied the power of shear experiments to discriminate between
them.  Fedeli et al. (2011) investigated the effect of baryon physics
on the matter power spectra for standard and extended quintessence
models using a box of 300 $\,h^{-1}$Mpc and $(2 \times) 768^3$
particles.  Viel et al. (2011) explored the effects of baryons of the
characteristic cut-off in the initial power predicted by the warm dark
matter model using a set of simulations of size ~6.25-100$\,
h^{-1}$Mpc and $(2 \times) 512^3$ particles.  We note also the study
of Guillet et al. (2010), who did not derive spectra but, almost
equivalently, studied the impact of gas dynamics on the variance (and
skewness) of the mass distribution in a box of 50$\, h^{-1}$Mpc with
$(2 \times) 1024^3$ particles.  Finally, Van Daalen et al. (2011)
  performed hydrodynamical simulations in boxes of various size, up to
  100$\, h^{-1}$Mpc on a side, by using $(2\times) 512^3$ particles
  and including AGN feedback, and used them to evaluate fluctuation spectra. In turn, Semboloni et al. (2011) used such spectra to derive their shear spectra.

Our simulations are run in a box of side $L = 410\, h^{-1}$ Mpc ($\,
$corresponding to $k_L \sim 1.5 \times 10^{-2}\, h$Mpc$^{-1}$, where
spectra are perfectly linear even today) and have a force resolution
$\epsilon = 7.5 \, h^{-1}$ kpc ($\, $corresponding to $k_\epsilon \sim
8.4 \times 10^{2}\, h$Mpc$^{-1}$), so that they can be used to produce data 
with high $k$ values. More precisely, spectra can be
generated up to $k \simeq N(2\pi /L)$
with $N \simeq 2^{15} = 32768$, while maintaining $k~( \simeq 5.0 \times
10^{2}\, h$Mpc$^{-1}) < k_\epsilon$.
A technique to obtain exact spectra up to such high--$k$, without
the use of exceedingly large grids, is described in Section 3.
The spectra we mostly use extends up to $k
\simeq 250\, h$Mpc$^{-1}.$

The plan of the paper is as follows. We present in Section 2 the
simulations on which our analysis is based. In Section 3, we first describe
the method to compute the power spectrum of density fluctuation
from simulations. After discussing the effect of numerical noise and
the comparing N--body results with those of {\em HF}, we then show
the effect of baryon physics on the power spectrum. In Section 4,
we pass from the fluctuation spectrum to the cosmic shear spectrum. We
discuss our results and draw our main conclusions in Section 5.  

\section{Simulations}
Our simulations follow the development of large-scale structure within
a periodic box of comoving size $L = 410\, h^{-1}$Mpc, using $(2
\times) 1024^3$ particles, in a spatially flat Gaussian $\Lambda$CDM
model with $\Omega_m = 0.24$, $\Omega_b= 4.13 \times 10^{-2}$,
$h=0.73$, and $n_s = 0.96$, (density parameters of total matter and
baryons, Hubble parameter, primordial spectral index, respectively)
and are normalized so that $\sigma_8 = 0.8$ at $z=0~.$ With this
parameter choice, our simulated cosmological model is consistent with the 
WMAP-7 CMB results (Komatsu et al. 2010). 

For the hydrodynamical simulations, we then have two populations
of particles, whose mass ratio is chosen to reproduce the cosmic
baryon fraction, initially placed on two uniform grids, displaced by 
half a grid size. Initial displacements from the
unperturbed grid position are been generated according to the
Zeldovich approximation at the initial redshift of $z_{in} = 41$. The
resulting masses of CDM and baryonic particles are $m_c \simeq 1.89
\times 10^9\, h^{-1} M_\odot$ and $m_b \simeq 3.93 \times 10^8\,
h^{-1} M_\odot$, respectively.

For the sake of comparison, we also run a purely gravitational simulation. 
In this case, the particles are initially set on a single
uniform grid and have a mass $m_c \simeq 2.28 \times 10^9\, h^{-1}
M_\odot$. They are then again displaced according to the Zeldovich
approximation to the initial redshift $z_{in}$.

Simulations are carried out using the TreePM smoothed--particle hydrodynamics
(SPH) GADGET-3 code, an
improved version of the GADGET-2 code (Springel 2005), which
allows each processor to also be assigned disjoint segments of the
Peano--Hilbert curve, on which the decomposition of the computational
domain is based. This provides a substantial improvement to the
balance among the work--loads assigned to the processors, thus
substantially increasing the code efficiency. Gravitational forces are computed
 using a Plummer--equivalent softening, which is fixed to
$\epsilon_{Pl}=7.5\, h^{-1}$ physical kpc from $z=0$ to $z=2$, and fixed
in comoving units at higher redshift.
The initial conditions are generated using the package
NgenIC, part of the GADGET distribution.

The simulation conteining only dark matter (DM) is dubbed DMo. 
We then have two hydrodynamical simulations.
One is a previously developed {\it non--radiative} simulation (gravitational
heating hereafter GH) that uses 64 neighbours to compute the hydrodynamic 
forces, with the width of the B-spline smoothing kernel
allowing us to reach a minimum value equal to half of the gravitational
softening. A second hydrodynamical simulation is carried out by
including the effect of cooling and star formation (hereafter CSF). In
this simulation, radiative cooling is computed for non--vanishing
metallicity according to Sutherland \& Dopita (1993), who also considered 
heating/cooling from a spatially uniform and evolving ultraviolet
background. Gas particles above a given threshold density are treated
as a multi-phase medium, so as to provide a sub–resolution description of the
inter–stellar medium according to the model described by Springel \&
Hernquist (2003). Within each multi-phase gas particle, a cold and a
hot-phase coexist in pressure equilibrium, with the cold phase
providing the reservoir of star formation. Conversion of collisional
gas particles into collisionless star particles proceeds in a
stochastic way, with gas particles spawning a maximum of two
generations of star particles. The CSF simulation also includes a
description of metal production from chemical enrichment by supernova (SN)
SNII, SNIa, and asymptotic giant branch stars, as described by Tornatore et
al. (2007). Stars of different mass, distributed according to a
Salpeter initial mass function (IMF), release metals over the timescale 
determined by the corresponding mass-dependent life times. Kinetic feedback is
implemented by mimicking galactic ejecta powered by SN explosions. In
these runs, galactic winds have a mass upload proportional to the
local star-formation rate. We use $v_w = 500\, {\rm km/s}$ for the
wind velocity, which corresponds to an assumption of about unity efficiency
for the conversion of energy released by SNII into kinetic energy for
a Salpeter IMF.  The same simulations were employed to analyse the 
contribution of baryon physics to the halo mass
function (Cui et al. 2011).

The feedback model included in the CSF simulation is known not to be
fully adequate for regulating overcooling, especially in large 
cluster--sized halos (e.g., Borgani et al. 2006). 
It is also known that including AGN feedback leads to a more
efficient regulation of star formation inside galaxy clusters.
In turn, over--cooling is certainly absent in non--radiative
simulations. Accordingly, CSF and NR simulations represent somehow two
extremes, and a fair description of physical feedback mechanisms
is expected to yield the results derived between them.

\section{Fluctuation spectra}

Power spectra are computed at 20 redshift values obtainable from the
expression $1+z_r = 10^{r/20}$ ($r=0,1,...,19$).  Spectra are
evaluated by using the algorithm PMpowerM included in the PM package
(Klypin and Holtzman 1997), courtesy of A. Klypin.  Through a CiC
procedure, the algorithm assigns the density field on a uniform
cartesian grid starting from the particle distribution. It then evaluates 
the spectrum by applying Fast Fourier Trasform (FFT) on a $n^3$ grid.

Here we consider the effective $n=2^fN$ values, with $f$ from 0 to 5,
i.e. $n$ from 1024 to $1024 \times 2^5 = 32768~.$ Such large $n$ are
obtainable by considering a $N^3$ grid in a box of side $L/2^f$, where
all simulation particles are inset, in points of coordinates $x_{i,f}
= x_i - \nu L/2^f$ ($i=1,2,3$), with an integer $\nu$ selected so that
$0 < x_{i,f} < L/2^f$. With this procedure, the size of the grid where
the density field is assigned reaches $\simeq 12.5\, h^{-1}$kpc.
This {\it folding} technique was first proposed by Jenkins et al
  (1996) and a wider description can be found in Smith et
  al.~(2003). In Appendix A, we briefly discuss why this tecnique is
  so effective.
The capability of this technique to yield spectra down to wavelengths
slightly above the gravitational softening scale is limited by 
the numerical noise of the grid used to set the initial
conditions. The problem, however, is not caused by the reductions in
the box size and would be identical if large $n$ grids could be
directly applied to the original box of side $L$.

\begin{figure}
\begin{center}
\includegraphics[height=9.cm,angle=0]{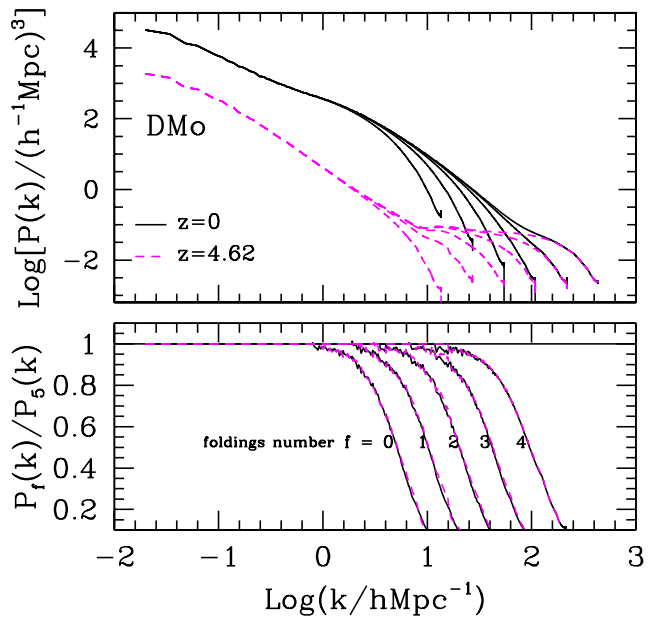}
\end{center}
\vskip -.5truecm
\caption{
  Upper panel: power spectra for the DM--only simulation, obtained by
  using $n^3$ grids with effective $n$ from 1024 to 32768, are shown
  at $z=0$ (solid curves) and $z=4.62$ (dashed curves). The latter
  is the highest redshift at which non--linear spectral features are
  (barely) visible above the numerical noise. In contrast, at $z=0$,
  numerical noise affects only the spectrum at the highest resolution. Lower
  panel: The ratios of $P_f(k)$ ($f$=0,.., 4) to $P_5(k)$ are
  plotted. Values of this ratio at the two considered redshifts are
  almost indistinguishable and this plot explicitly shows that they do
  not depend on the extension of numerical noise and are self--similar
  when the resolution limit is shifted.  }
\label{dm}
\end{figure}

\subsection{Dark matter simulation: spectral extension and
  numerical noise} 
In Figure \ref{dm} (upper panel), we overlap the spectra $P_f(k)$ of
the DMo simulation at $z_0=0$ and $z_1=4.62$, obtained for values of
the folding parameter $f$ ranging from 0 to 5, i.e. by using an
effective $n^3$ grid, with $n$ ranging from 1024 to $2^5\times
1024$. At low $k$, baryonic acoustic oscillations (BAOs) are visible,
with the expected amplitude, that are slightly attenuated 
from $z_1$ to~$z_0$.

While the onset of non--linearity is evident in the spectrum at $z=0$,
the situation is quite different at the higher redshift. Here
numerical noise affects $k$ values even below $\sim 10$, almost
completely precluding the analysis of nonlinearity, displaying its
main effects above such scale, at $z_1~$.
\begin{figure}
\begin{center}
\includegraphics[height=7.5cm,angle=0]{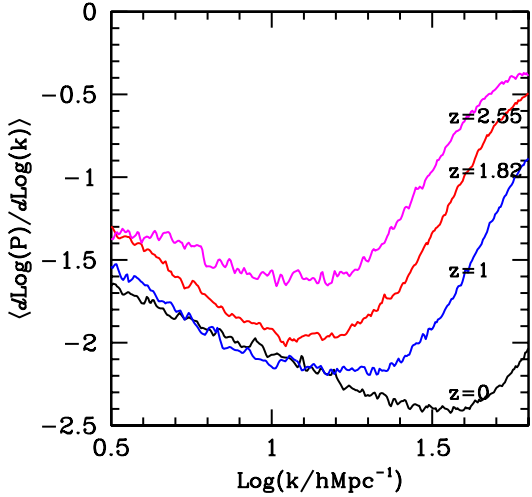}
\end{center}
\vskip -.5truecm
\caption{Smoothed (see text) logarithmic derivatives of $P(k)$ at
  various $z$ values (here $f=5$).
}
\label{deriv}
\end{figure}

We note however that the shapes of the smearing of the $P_f$ spectra
($f=0~$...~$ 4$), with respect to $P_5$, are unaffected by numerical
noise. In the lower panel the cuts for $z_0$ and $z_1$ are indeed 
both plotted, but they can hardly be distinguished. In the lower panel
of Figure \ref{dm}, one can also appreciate that the smearing shape is
similar for any $f$ value, just logarithmically displaced either up-- or
downwards.

These results highlight that two criteria that need to be considered when
making an optimal choice of the number $f$ of foldings, 
which is large enough to allow us to make use
of spectra at large $k$ values. At the same time, one should 
consider the onset of numerical noise. In general, it is futile to attempt 
to derive spectra in regions where the numerical noise dominates.

We then consider, for instance, the case $n=N$ ($f=0$). The formal
comoving resolution length is then $L_{res} = L/N \simeq 0.40\,
h^{-1}$Mpc, corresponding to $k_{res} = 2\pi/L_{res} \simeq 15.7\,h$
Mpc$^{-1}$. The lower panel of Figure~\ref{dm} however shows an almost
vanishing spectrum at this wavenumber. As a matter of fact, the
spectrum is already smeared by a factor $\sim 0.8$ at $k \simeq
3$. Accordingly, quite in general, spectra are unreliable when $k$
increases above 20$\, \%$ of the formal resolution limit $k_{res}~.$
In this case, clearly, $f$ ought to be suitably increased. We then need
to ensure that no shot noise will interfere with this.

As a matter of fact, even a random distribution of $N$ particles
exhibits a non--vanishing spectrum $\propto N^{-1}$. Smith et
al.~(2003) performed a detailed analysis of this numerical noise, by
studying the evolution of 236$^3$ particles initially distributed with
a spectrum $\propto k^{-2}$. 


Here, we define the scale where the numerical noise begins to affect
spectra by evaluating a numerical derivative $\partial P(k,z)/\partial
k$ and fixing it to the $k$ values where the spectrum turns from
decreasing to increasing. This evaluation requires a suitable
smoothing of the spectra (five--point Gaussian averaging) and the
derivative itself includes an additional smearing (again of five--point 
Gaussian averaging) to appear as shown in Figure \ref{deriv}.
Taking into account the different features of our simulations,
our results are consistent with those of Smith et al.~(2003).

The behaviour of $\partial P(k,z) /\partial k$ shown in Figure 2
then indicates a limitation of the use of spectra shifting from 
$k \simeq 40$--50$~h\, $Mpc$^{-1}$ (comoving $L /h^{-1}{\rm Mpc}
\simeq 0.09$) at $z=0$, to 10--12 (0.42--0.52) at $z=2.55~.$ As we
show, the contributions to shear observables from higher redshift
become quite small. Accordingly, at $z=0$ the numerical noise affects
$k$ values approximately down to $k_\epsilon/5~$. The $P_4(k,z=0)$
spectrum is then substantially unaffected by numerical noise, which
still causes some distorsion in $P_5$, and limits the
use of spectra at $z=0$.

A key issue that we anticipate is that shear spectra, resulting from
an integration along $z$ over variable $k$ values, require the
knowledge of $P(k,z)$ at progressively smaller $k$ as $z$
increases. Since the integration involves passing from a
spatial to an angular spectrum, the same angle $\vartheta \sim [$$\cal
O$$(2\pi/\ell)]$ subtends an increasingly larger scale $\lambda \sim [$$\cal
O$$(2\pi/k)]$ at greater distances. This effect occurs approximately
in parallel with the downward shift in $k$ caused by the impact of numerical
noise. Thus, the spectral range we need to consider at higher $z$ is
mostly noise free, once it is so at $z=0$.

We however emphasize that, in recent works dealing with the effects
of baryons on the power spectra, no $n$ value above $N$ was used to
produce spectra.
With a box of $\sim 60\, h^{-1}$Mpc and $N=512$, as used by Rudd et
al. (2008), we then have $L_{res} \simeq 0.12$ and $k_{res} \sim 50$.
However, as shown in the lower panel of Figure \ref{dm}, spectral
information is scarce above $k \sim 10 \simeq k_{res}/5\, $, even
at $z=0~.$ This place severe limitations on our ability to evaluate 
$P_{ij}(\ell)$ at large $\ell~.$ In our case, a similar limit is attained 
only at $z \sim 2.5$, while
at lower $z$ values, fluctuation spectra are reliable well beyond which
$k$ value.

\begin{figure}
\vskip -.4truecm
\begin{center}
\includegraphics[height=9.5cm,angle=0]{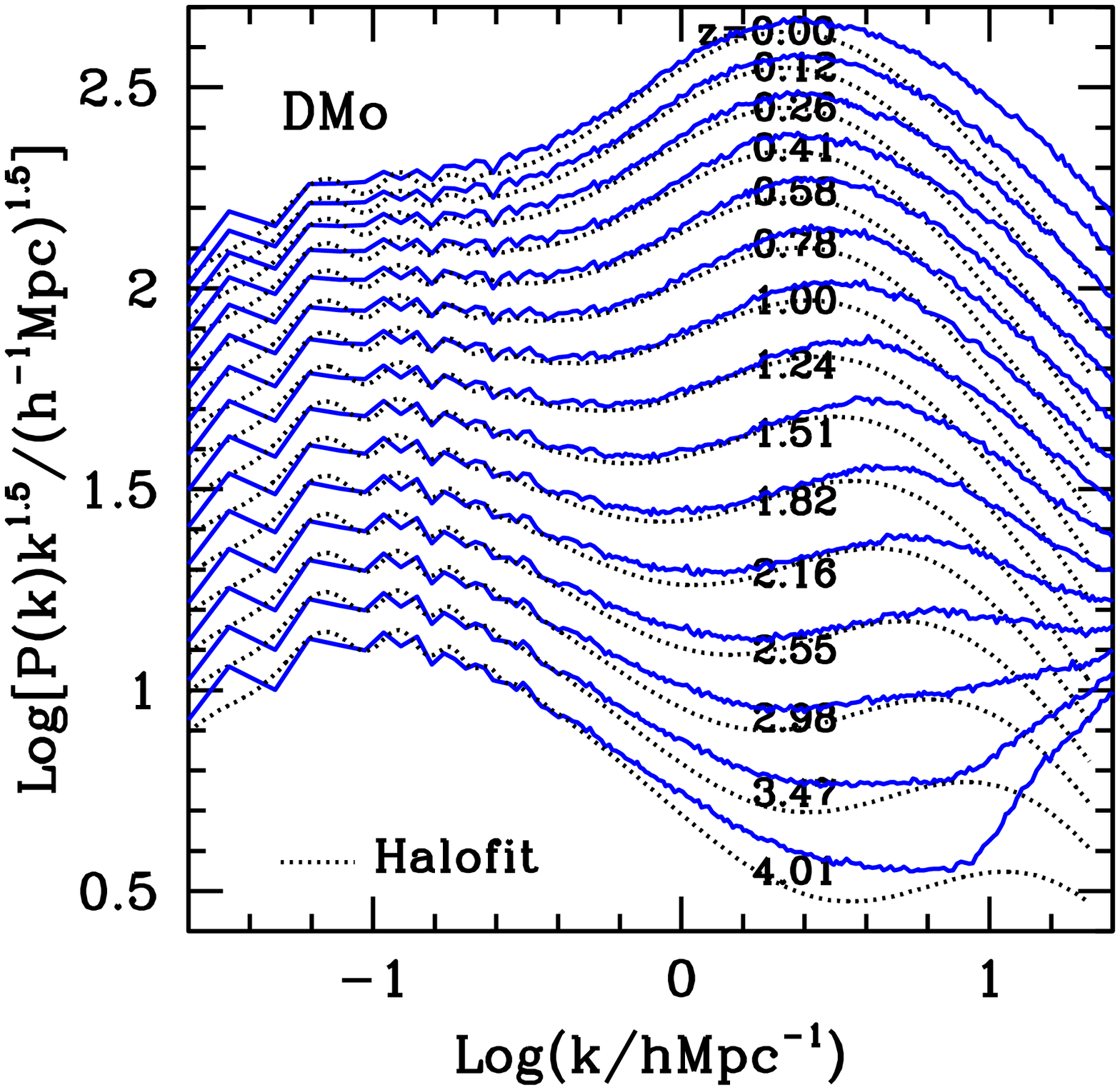}
\vskip -.3truecm
\includegraphics[height=9.5cm,angle=0]{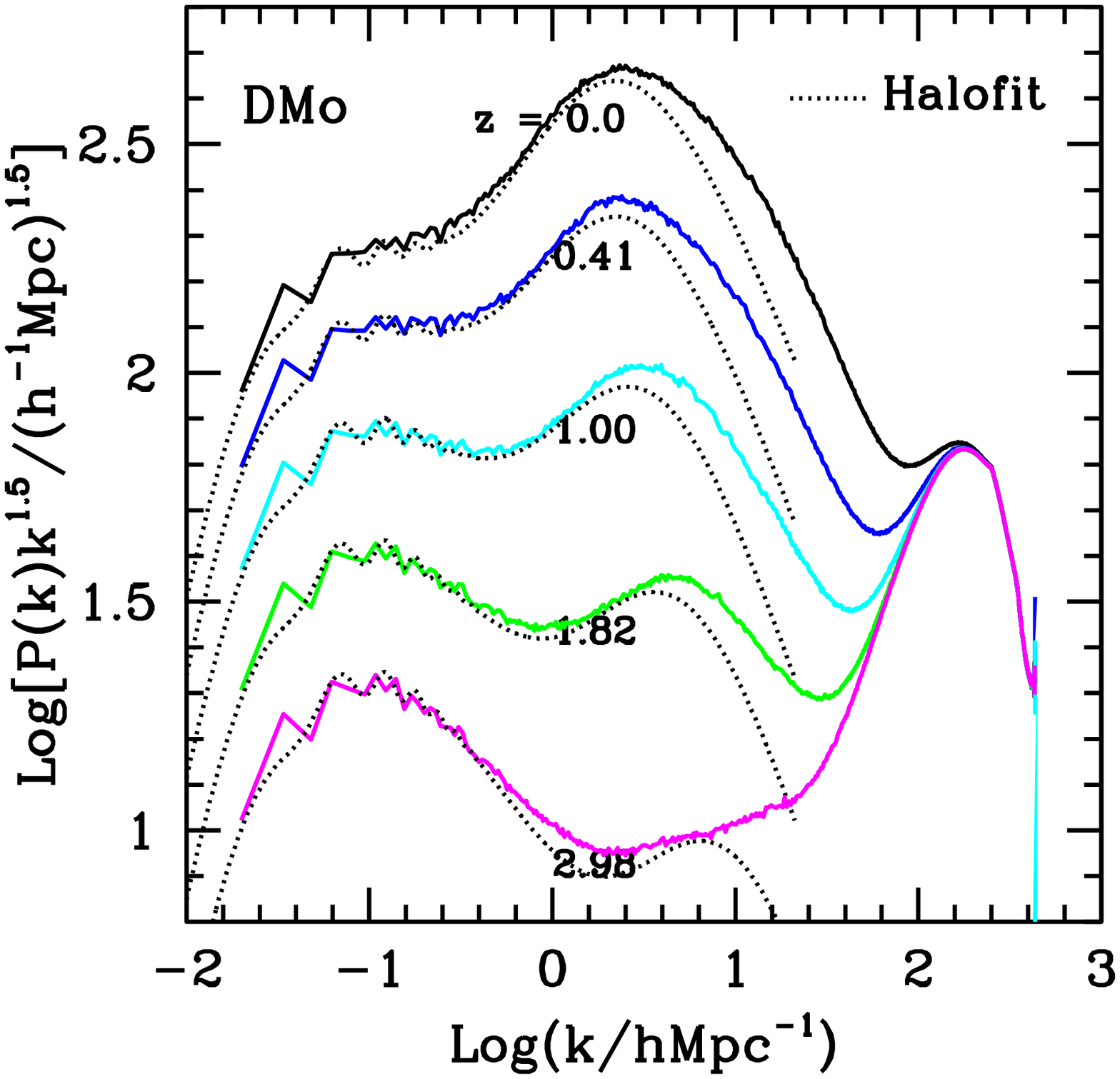}
\end{center}
\vskip -.5truecm
\caption{Spectral evolution, for DMo simulations, in the redshift
  range used to compile shear spectra. Spectra obtained with $f=5$ are
  shown. Simulations are also compared with the {\it HF}
  predictions. In the top panel, we show all spectra (multiplied by
  $k^{3/2}$, to allow an overall comparison) in a restricted $k$
  range. Redshift values are shown along aside each curve. In the bottom
  panel, we select five redshift values and plot spectra over a wider $k$
  interval. Substantial discrepancies from HF are evident, even at low
  $k$ values, at any redshift. }
\label{hf}
\vskip -.3truecm
\end{figure}

\subsection{Dark matter simulation: Simulation spectra versus Halofit}
In Figure \ref{hf}, we explore the spectral evolution across the redshift
range $z=0$--4~, by showing $k^{3/2} P(k,z)$. Power spectra for the
DMo simulations (solid curves) are obtained with $f=5$ and are also
compared with {\it HF} predictions (dotted curves).

In the top panel, we give an overall picture of all spectra used up to
$z \sim 4$, in a restricted $k$ range. In the bottom panel, we widen
the $k$ range from 0 to 3, but restrict ourselves to five spectra.

These plots illustrate that there is a considerable disagreement between 
the HF predictions and, those of the simulations, significantly exceeding 
the $\pm 3$ per cent approximation
claimed in Smith et al. (2002)\footnote{ Smith et al (2002), however, 
claimed to reproduce the results of their simulation with HF only
in a restricted range of $k$ and $z$, and not in general }. 

At $z \sim 3$--4, in the range $k \sim 1$--2$\, h\, $Mpc$^{-1}$, we
find discrepancies exceeding 15--20$\, \%$.  In general, we note that
there is a systematic lack of power in HF, with respect to
simulations, even in the region of a non--linearity onset.  Such a
lack of power bursts above $k \simeq 10\, h$Mpc$^{-1}$.  As we discuss
below, this is however the regime where the baryon physics should be
most important and the predictions of the DMo simulation less
reliable. When AGN feedback is included, Van Daalen et al. (2011)
  also noticed that baryon physics affects the matter distribution
  down to $k \sim 0.3$--$1\, h\,$Mpc$^{-1}$.

A robust analysis should certainly be based on more model realizations. 
We however recall that a lack of power would be the natural
consequence of a shortage of long--wave contributions. The HF expressions,
which have been calibrated using simulations within boxes of sizes
never exceeding 240$\, h^{-1}$Mpc (see Jenkins et al. 1998, Smith et al.
2003), could actually lack some power, leading to a delayed onset of
non--linearity.

\subsection{Hydrodynamical simulations: total spectra}
We then consider the effects of baryon physics, starting from the
analysis of
the non--radiative (GH) simulation.

\begin{figure}
\begin{center}
\includegraphics[height=9.cm,angle=0]{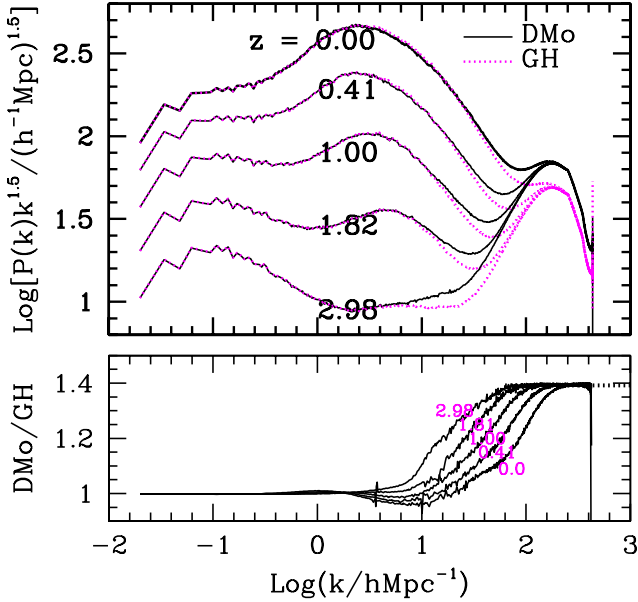}
\end{center}
\vskip -.5truecm
\caption{Power spectra of density fluctuations for the non--radiative
  (GH) hydrodynamical and DMo simulations. Results are shown for
  $f=5$. Top panel: evolution between $z \simeq 3$ and $z=0$ for the
  GH (solid curves) and DMo (dotted curves) simulations. Redshift
  values are indicated at the side of each curve. Here, the small spectral
  differences are hardly visible, up to the onset of white noise at
  large--$k$. Bottom panel: ratio of power spectra of the DMo to
  GH simulations. A slight excess of power in the GH simulation is
  visible up to $k \simeq 10\,
  h$Mpc$^{-1}$. The inversion at higher $k$ value is
  covered by the onset of white noise (see also Figure
  \ref{deriv}). Note the different levels of white noise in the two
  simulations, owing to the different numbers of particles used (see
  text). }
\label{svisc}
\end{figure}

\begin{figure}
\begin{center}
\includegraphics[height=9.cm,angle=0]{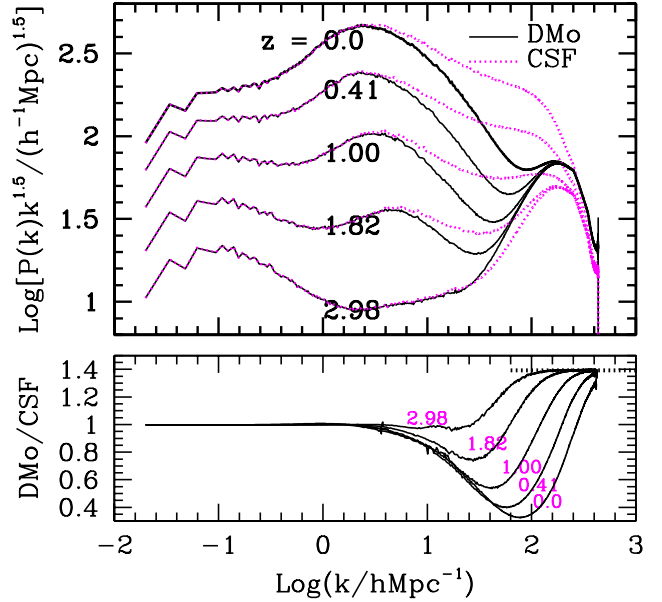}
\end{center}
\vskip -.5truecm
\caption{The same as in F \ref{svisc} but for the comparison between
  the radiative (CSF) simulation and the DMo simulation.  Top panel:
  the larger amplitude of the CSF power spectrum is clearly visible,
  well before the onset of white noise at large--$k$. Bottom panel:
  ratio of the DMo to CSF spectra. A slight lack of power in CSF, up
  to $k \simeq 10\,
  h$Mpc$^{-1}$, is hardly appreciable.}
\label{csf}
\end{figure}

In Figure \ref{svisc}, the total fluctuation spectra of the GH
simulation are compared to DMo spectra. The pressure support of the
baryonic component partly inhibits the increase in the fluctuations in the
total matter distribution, thereby reducing the amplitude of the power
spectrum on scales below the corresponding Jeans length. 
In Figure \ref{csf}, we present the same comparison for the CSF
simulation, which includes the effect of radiative cooling and star
formation.

In the GH case, 
the reduction in the power spectrum amplitude due to
baryons gradually moves towards larger $k$ values as one moves to lower
redshift. The onset of white noise can obscure this effect, which is
hardly visible above $z \sim 1~.$ The average
interparticle separation is different in the DMo and GH simulations, since
two particle populations (one collisionless the other collisional),
displaced by half grid cell, are used to sample density and velocity
fields in the initial conditions of the hydrodynamical
simulations. Therefore, twice as small as number of particles in
the DMo simulation justifies the correspondingly higher level of white
noise reached above the Nyquist frequency. This effect must be kept
in mind in any comparison between DMo and hydrodynamical simulations.

Owing mostly to the larger range covered by the
$y$-axis, it is even harder to appreciate here a slight lack of power
in the CSF simulation at $k \sim 1\, h$Mpc$^{-1}$. 
The spectra for the CSF
simulation are then systematically above those of the
DM--only case. This increase in the small--scale power in the CSF case is due
to the contraction of halos induced by gas overcooling (see also Rudd
et al. 2008), which alse removes the pressure support responsible for the 
inhibition of the fluctuaction growth in the GH case.

\begin{figure}
\begin{center}
\vskip -1.42truecm
\includegraphics[height=9.cm,angle=0]{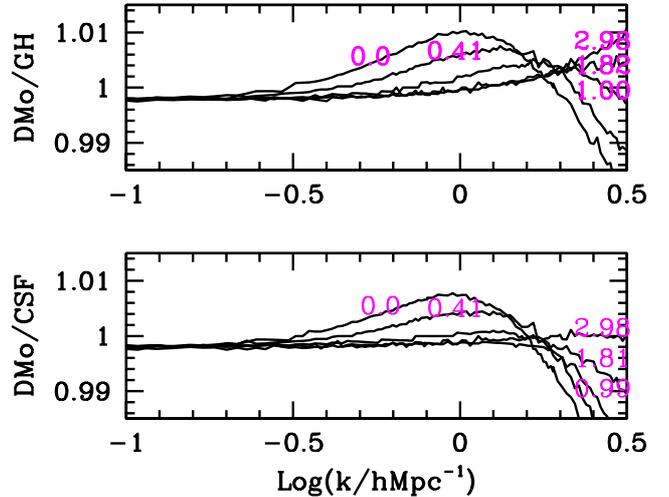}
\end{center}
\vskip -.5truecm
\caption{A blown-up view of the lower panels of Figures \ref{svisc},
  \ref{csf}, to show the slight inversion of the ratio of the power
  spectra of DMo to hydrodynamical simulations, which takes place
  around $k \sim 1\, h$Mpc$^{-1}$; an effect previously found by
  various authors (see text). The effect is smaller in CSF than in
  GH. }
\label{magnify}
\end{figure}

Quite interestingly, in both GH and CSF simulations, 
we found the inverse effect around $k \sim 1\, h\,
$Mpc$^{-1}$, where a slight increase in the spectral amplitude appears.
This effect is more apparent in the
upper panel of Figure~\ref{magnify}, which provides a zoomed view.


The slight excess power around $k \sim 1\, h$Mpc$^{-1}$
in the DMo simulation has also been by analyses of
the power spectrum in cosmological N--body and hydrodynamical
simulations ( Van Daalen et al. 2011 \footnote{In this paper we compare our CSF simulation with the REF one in Van Daalen et al 2011}). This
feature occurs in a dynamical range where the effects of
non--linearity start to appear.
A similar effect was also outlined by Rudd et al. (2008), also
using both a non--radiative and a radiative simulation, similar to our
GH and CSF cases, but carried out with the adaptive mesh refinement (ART)
code (Kravtsov et al. 1997). In addition Casarini et al.~(2011b) found a
similar effect in hydrodynamical simulations carried out with the
SPH {\sc Gasoline} code (Wadsley et al. 2004), using a box of 64$\,
h^{-1}$Mpc on a side.

The size of the effect found here can be appreciated from
Figure~\ref{magnify}. While Rudd et al. (2008) found an effect as
large as ten per cent, Van Daalen et al. (2011) and Casarini et
al. (2011b) found a smaller effect, ranging from one to two percent, thus
closer to the one obtained from our analysis.  Furthermore,
differences also exist in the $k$ range where the effect arises. It is
quite difficult to draw a strong conclusion from this comparison
about the analyses based on simulations covering different dynamic
ranges, and using different implementations of radiative cooling and
star formation. Nevertheless, we note that the
largest feature of power inversion appears in the simulation carried
out with a Eulerian grid--based code, with respect to the other
SPH--based analyses.

\subsection{Hydrodynamical simulations: behaviour of the different
  components} 
\begin{figure}
\begin{center}
\includegraphics[height=8.6cm,angle=0]{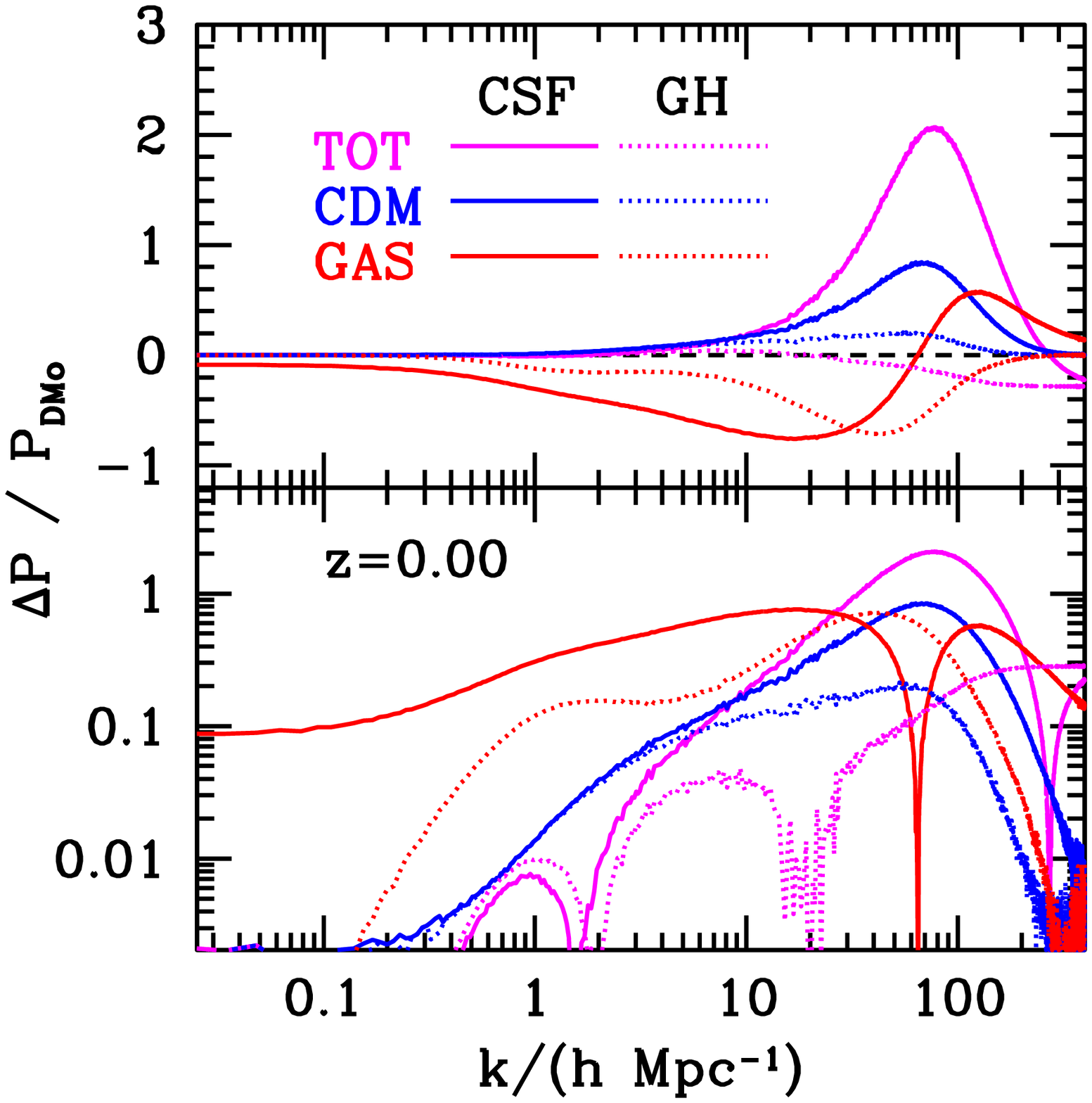}
\vskip -0.8 truecm
\includegraphics[height=8.6cm,angle=0]{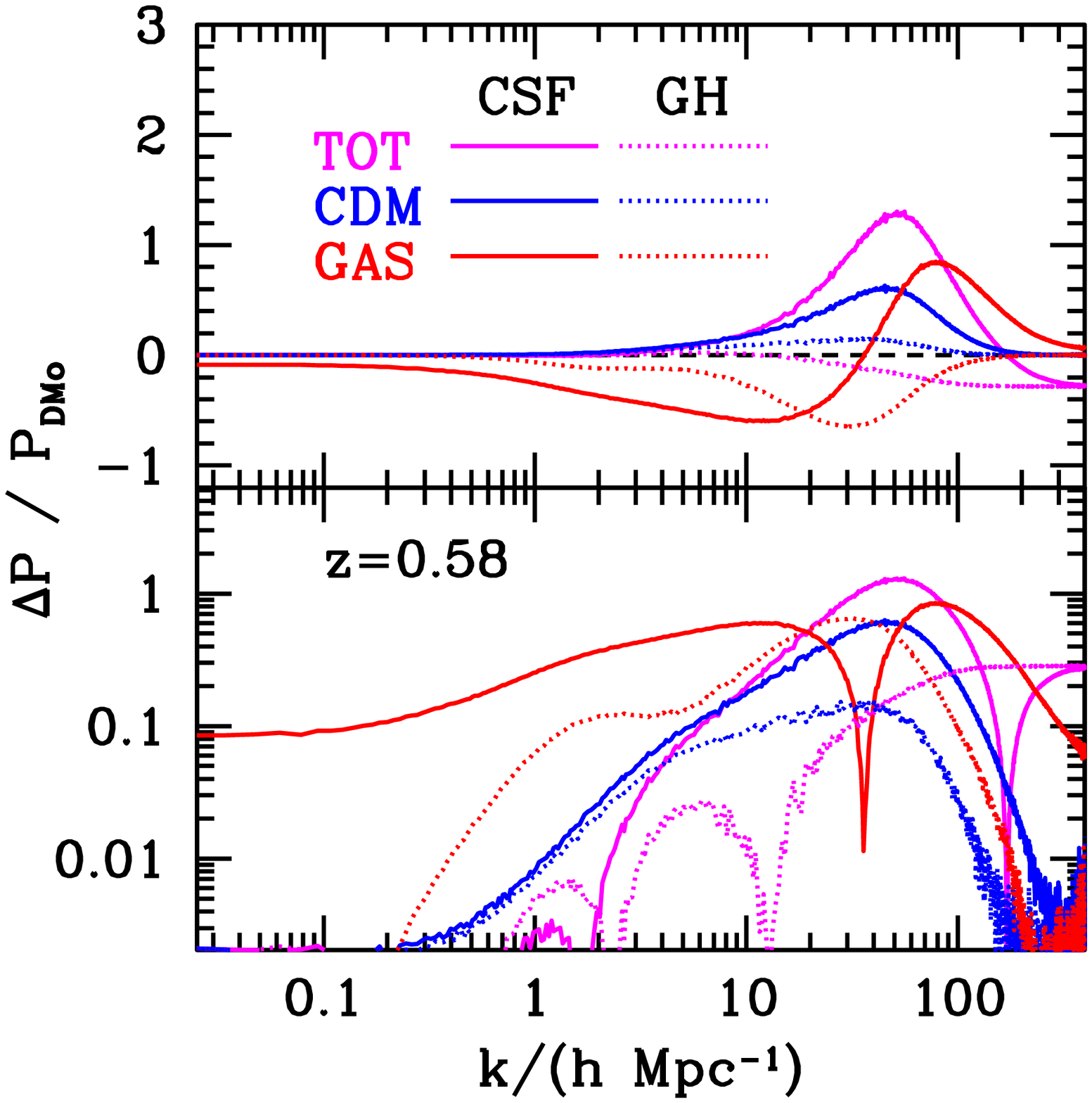}
\vskip -0.8 truecm
\includegraphics[height=8.6cm,angle=0]{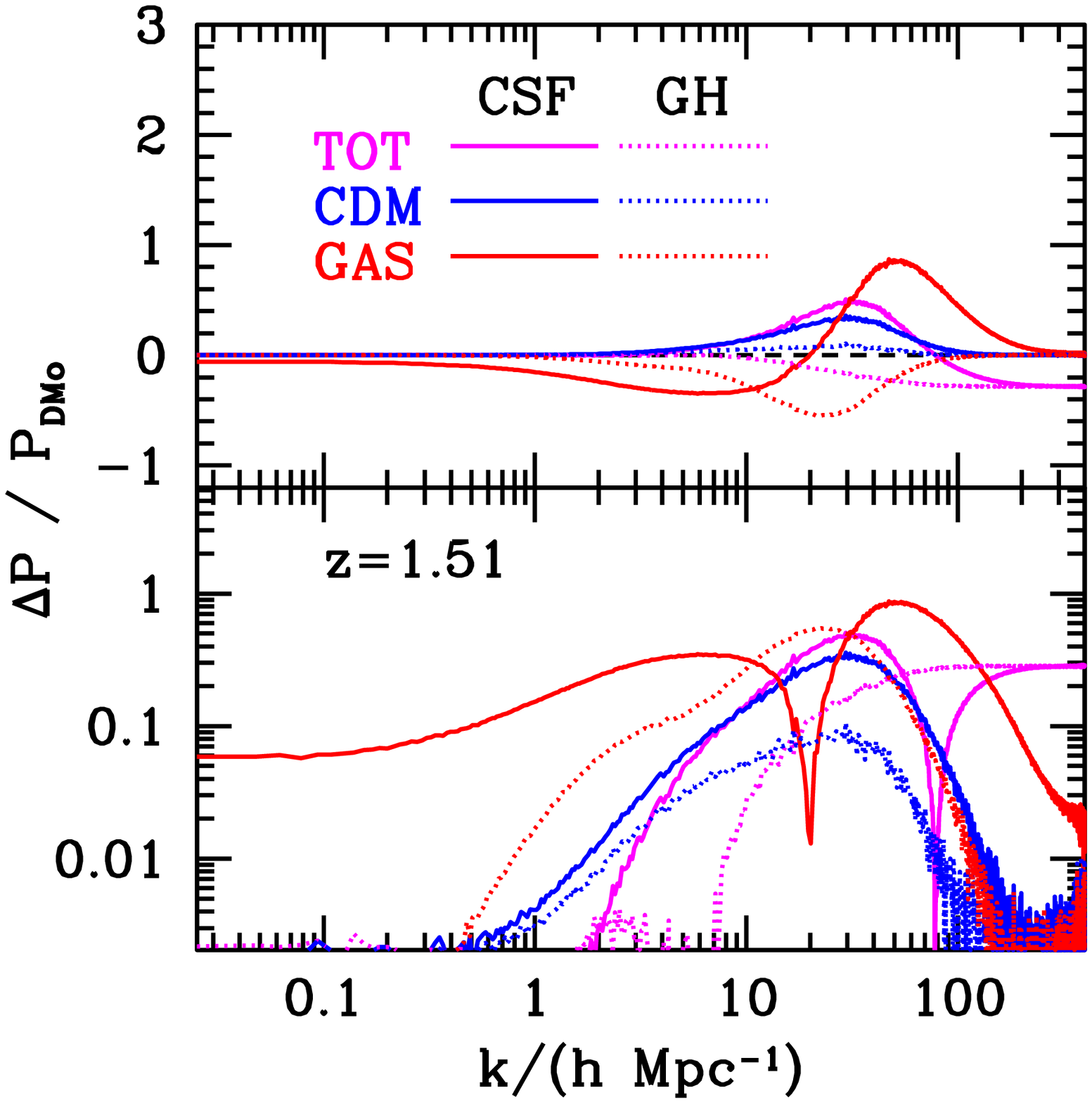}
\end{center}
\vskip -0.8 truecm
\caption{Spectral evolution of the different components in the GH and CSF
  simulations. For comments, see text.}
\label{c1}
\end{figure}
In Figure \ref{c1}, we show the power spectra of the DM and gas components in
the hydrodynamical simulations, normalized to the corresponding power
spectrum computed for the DMo case. The three panels show results at $z =
0$, 0.58, 1.51 (from top to bottom). The upper and lower parts of each
panel show results plotted using linear and logarithmic units on the
$y$-axis, respectively.

We outline a few points that are directly observable there.

\noindent
First of all in the GH case, the total spectrum is mostly above the
corresponding spectrum of the DM component, while the gas spectrum
is increasingly below that of the DMo simulation.

In contrast, in the CSF case the total spectrum exceeds that of the DM,
 above $k \sim 9$--10$\, h$Mpc$^{-1}$. This is presumably due to
the contribution of stars, which are highly concentrated within halos,
thus providing a large amount of fluctuation power on small scales.
The same feature, although appearing at slightly different values of
$k$, was also found by Rudd et al. (2008), Van Daalen et
al. (2011), and Fedeli et al. (2011).

Furthermore, as we have already pointed out, hydrodynamics acts as a brake
on the growth of overdensities on scales smaller than the Jeans
length. Therefore, in the non--radiative GH case the fluctuation
amplitude of the gas component becomes smaller than that of the DM
component.  

The presence of radiative cooling allows gas to
dissipate energy and sink to the centre of DM halos, thereby
increasing their concentration by adiabatic contraction
(e.g. Gnedin et al. 2004, Sellwood \& MacGough 2005). At $z=0$, this
causes an increase in the fluctuation power on small scales $k> \sim
40\, h$Mpc$^{-1}$, namely on a length scale $\sim 150\,
h^{-1}$kpc, close to the scale length where CDM and gas density
profiles begin to diverge from each other.

While, in Van Dalen et al. (2011), the total spectra were found to 
overcome CDM around
the same $k$ as here, the scale where the gas spectrum overcomes the
DMo spectrum is greater.

\noindent
Secondly ,the evolution in redshift of the total power spectrum and its
components allows us, first of all, to follow the progressive onset of
the inversion in the ratio of DMo to either CSF or GH spectra, as
described in section 3.3~. Logarithmic plots are appropiate way of following
this process.

At redshift as high as $z=1.51$, when the $k$--range below 10$\, h\, $Mpc$^{-1}$
is clearly distinguishable above numerical noise, we can discerne an
inversion bump at $k \sim 2$, in the GH total spectrum, whose
amplitude attains $\sim 0.5\, \%$. At $z=0.58$, the inversion interval
has widened in GH, reaching an amplitude $\sim 1\, \%$, and is also
visible in the CSF case. Altogether, however, the evolution of
the GH/DMo ratio exhibits a waving development, with a power deficit
at short wavelengths. In CSF, the inversion bump is instead 
isolated and does not exceed $1\, \%$, even at $z = 0~.$

\noindent
Our third remark is that there are discrepancies between our findings and Rudd et al (2008). These authors, found a magnification of the inversion
in the presence of cooling and star formation. Furthermore, they were unable
to distinguish the ``waving'' behaviour in the GH case, in the
counter--inversion interval when $k$ is between $\sim 2$ and 15. On 
the other hand, our finding confirms Casarini et al. (2011b), Van
Daalen et al. (2011), and Fedeli et al. (2011) results. These 
discrepancies might arise from the different codes used or
from the too small box used in Rudd et al. (2008).

\noindent
Our fourth is that the high--$k$ behaviour at high $z$ allows us to appreciate
 in detail how the different level of numerical noise arises in the
different components. In particular, the amounts of power for CDM and
GAS in either GH or CSF coincide with that for DMo, because each component
share the size of the grid where initial conditions were set, even if particles
are displaced by half a cell. In the hydro cases, the
total spectrum noise exceeds the DMo one by the known factor.

\noindent
The fifth remark is that when cooling is allowed, we clearly note 
that gas loses energy more rapidly than CDM, and 
it then falls towards the centres of the forming halos. 
It is mostly because of gas accumulation on small
scales that the total power increases at large k values.  It is also
interesting to note that this process scales down with redshift.
This could partly be caused by the increase in the concentrations in
halo profiles, which is also by the presence of stars. However, we also 
note that baryon physics has characteristic length and time scales that are 
approximately constant in physical space, 
while $k$ is comoving and thus increases with decreasing redshift. 

We finally note that it is indeed significant that the results
of Casarini et al. (2011b), in spite of the different resolution, box
size, SNa cooling and numerical code, are consistent with those
found here.

\subsection{Connecting linear and simulation spectra}
To construct fairly well-normalized shear spectra, density
fluctuation spectra must extend down to $k \sim 0.01\, h$Mpc$^{-1}$,
or even less, in the full linear regime. For instance, some window
functions $W_i(z)$ imply that there is a significant contribution from
 power spectra to shear spectra at $z \sim 1$--1.5. 
The typical wave numbers for which fluctuation spectra are known are then
\begin{equation}
k \simeq \ell/[\tau_o-\tau(z)]~.
\end{equation}
Where $\tau$ is the conformal time in the metric
\begin{equation}
ds^2 = a^2(\tau)(d\tau^2 - d\eta^2)~,
\end{equation}
$\eta$ and $a(\tau)$ being the elementary space interval and the
scale factor, respectively, and $\tau_o$ and $\tau(z)$ are the
conformal age of the Universe and its conformal age at redshift
z~. Accordingly, $\tau_o - \tau(z) \sim 3000~h^{-1}$Mpc and $\ell \sim
30$ yields $k \sim 10^{-2}\, h$Mpc$^{-1}$.

In turn, this implies that using simulations within boxes of
$L \sim 60\, h^{-1}$Mpc on a side (i.e. reaching $k=2\pi/L\sim 0.1\,
h$Mpc$^{-1}$), no direct connection between linear and simulation
spectra can be made. In this case, one has then to resort to
non--linear approximations, such as {\it HF}, to cover an intermediate
$k$--range, up to a $k$ value where simulations provide a sufficiently
large number of Fourier modes (for $L=60\, h^{-1}$Mpc,   
discreteness effects in the sampling of these modes are significant 
even at $k \sim 0.3$--$0.5\, h$Mpc$^{-1}$). 
Delicate normalization problems then often have to be
solved. Although sharing the same {\it linear} $\sigma_8$,
DMo and HF spectra differ at most $k$ values, as shown in
Figure 3. Therefore, when a too small simulation box is used, the
discrepancy often has an opposite sign, and exhibits a different size
for different cosmologies.

The alternative option of comparing simulations with equal values of {\it
  non--linear} $\sigma_8$ improves the situation, since it overcomes
at least the part of the discrepancy arising from the different timing
of growth. This however requires a complex ``trial and error''
procedure, which clearly becomes less and less demanding as the box
size is increased. For instance, Casarini et al. (2011a), by using $L
\simeq 250\, h^{-1}$Mpc, reach a satisfactory convergence with only a
couple of iterations.

Directly connecting the power spectra from simulations and linear
theory eliminates any such problem. Although, at small $k$, simulation
spectra exhibit a significant discreteness, our box size of
410$\, h^{-1}$Mpc provides a connection between simulation and linear
spectra in a $k$ range where a sparse sampling of $k$--modes induces
only marginal effects. Following the same procedure as in Casarini et
al. (2011a), we slightly smooth discreteness effects above $k \simeq
0.1$ by averaging each $P(k_n)$ value with $P(k_{n-1})$ and
$P(k_{n+1})$,
giving $k_{n-1,n,n+1}^\alpha P(k_{n-1,n,n+1})$ ($\alpha$ selected so that
$k_{n-1}^\alpha P(k_{n-1})$ and $k_{n+1}^\alpha P(k_{n+1})$ are equal)
weight 0.15, 0.7, 0.15~, respectively.
We verified that small variations of these weights have no visible
impact on our final results.

\begin{figure}
\begin{center}
\includegraphics[height=9.cm,angle=0]{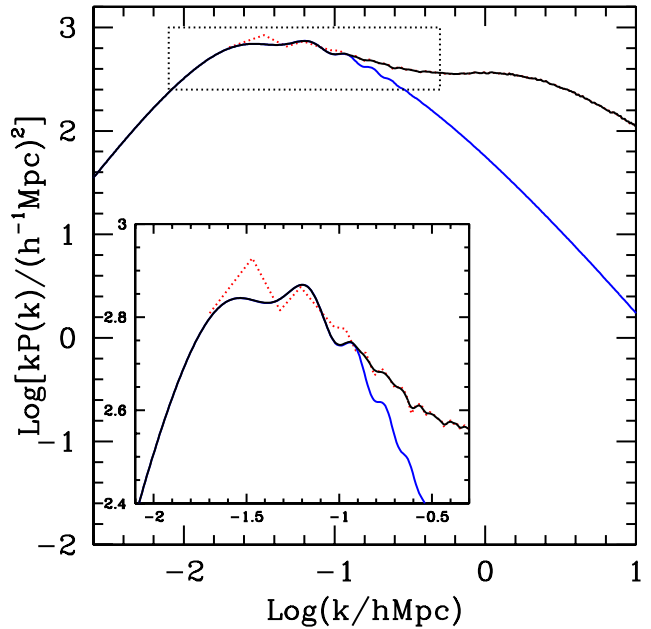}
\end{center}
\vskip -.5truecm
\caption{Connecting linear and simulation spectra for the
  hydrodynamical radiative (CSF) simulation at $z=0$. The line colours
  are as follows: blue is the linear spectrum; dotted red is the
  simulation spectrum; black is the connecting spectrum. The junction
  area (marked with the dottet rectangle) is magnified in the inset
  box, to also show the (mild) effect of the three-point smoothing on the
  simulation spectrum. Starting from large $k$, we pass from the
  simulation spectrum to the linear spectrum, where the former one
  approaches the latter for the first time.}
\label{connect}
\end{figure}
In Figure \ref{connect}, we show the connection between the linear and
the simulation spectra for the CSF simulation at $z=0$. The linear spectrum
shown here is the same as that used to generate our initial conditions of the
simulation.

We note here that the spectrum produced by the simulation
still exhibits the expected BAOs at the correct $k$ values. This is
true even where discreteness effects are still dominant. A mild
non--linearity is already visible above $k \simeq 0.1\, h\,
$Mpc$^{-1}$. The BAOs expected up to $k \sim 0.4\, h\, $Mpc$^{-1}$
can also however be traced in the simulation spectrum, despite the small 
amount of noise arising from sample variance. Such a small amount of 
contamination of the signal can easily be damped by running simulations of
different realizations of the same model in a similar box and averaging
among them.

\section{From fluctuation to shear spectra}
\subsection{Theory}
The angular power spectra for the weak lensing convergence are given
by the convolution of the matter power spectrum with window functions
that also account for the redshift distribution of the population of
lensed galaxies. Following Hu (1999),
if we bin the galaxies into $n$ redshift intervals, the convergence
power spectrum between the $i$ and $j$ tomographic beams, covering the
redshift intervals $\Delta_i$ and $\Delta_j$, is given by
\begin{equation}
P_{ij}(\ell) = H_{0}^{4}
\int_{0}^{\tau_0} du~ W_{i}(u) W_{j}(u)~ {P \left({\ell \over u},u
  \right)} ~.
\label{pijl}
\end{equation}
Here $H_0$ is the Hubble constant at present time, where $h$ is 
in units of $100\,{\rm km\,s^{-1}\,Mpc^{-1}}$, $\tau_0$ is the
  conformal time corresponding to present cosmic age, and $P(k,u)$ is the
  fluctuation spectrum at the conformal time $\tau = \tau_0 - u\, \, $.
  This angular convergence spectrum exhibits no explicit dependence on
  $h$, while it depends only indirectly on the cosmic expansion
  history through the $W_i(u)$ window functions. In the literature 
(Hu 1999, Casarini et al. 2011a),
$n=1$, 3, or 5 bins have been considered.  Here we mostly refer to a
five--bin case and the bin limits $z_i$ are selected so as to ensure that there is the same number of galaxies per bin.

We then define the window functions $W_i(u)$, which tell us how
clustered matter acts on the galaxies in the $i$--th bin. We
assume that the number density of galaxies per unit redshift interval
and solid angle is given by
\begin{equation}
n(z) = {d^2 N \over d\Omega\, dz} = {\cal C} ~\bigg({z \over z_0
}\bigg)^A \exp\bigg[- \left( z \over z_0 \right)^B \bigg]
\end{equation}
with
\begin{equation}
{\cal C} = {B \over \left[z_0 \Gamma \left( A+1 \over B
\right) \right]}~.
\label{nz}
\end{equation}
Here we take the usual values of $A=2$ and $B=1.5$, so that ${\cal C}
= 1.5/z_0$ where $z_m = 0.9~$ and $z_0 = z_m/1.412$ ($z_m$ median
redshift; see, e.g., Refregier et al. 2006, La Vacca \& Colombo 2008
and references therein).

This distribution is then considered within the limits of the redshift
bins.  
Following Hu et al. (2006), we account for the discrepancies between 
the photometric
redshifts, on which the redshift distribution is based, and the true
redshifts, by defining the filters
$$ 
\Pi_i (z) = \int_{z_{ph,i}}^{z_{ph,i+1}} dz' ~ {1 \over \sqrt{2 \pi}~
\sigma(z)} \exp \left(-{(z - z')^2 \over 2 \sigma^2 (z)} \right) =
~~~~~~~~~~~~
$$
\begin{equation}
~~~~~~~~ = {1 \over 2}
\left[
{\rm Erf}\left(z_{ph,i+1}-z \over \sqrt{2}\sigma(z)
\right)-{\rm Erf}\left(z_{ph,i}-z \over \sqrt{2}\sigma(z)\right)
\right],
\end{equation}
where $\sigma(z) = 0.05~(1+z)$ (see, e.g., Amara \& Refregier 2007, for
the motivation of this parameter choice).
We point out that this expression for the filters assumes a Gaussian
distribution for the errors in photo--$z$'s. An accurate
calibration of this error distribution is clearly of paramount importance to
avoid introducing biases into the reconstruction of power spectra derived from
weak lensing tomography (e.g. see Bernstein \& Ma 2008).

\begin{figure}
\begin{center}
\includegraphics[scale=0.42]{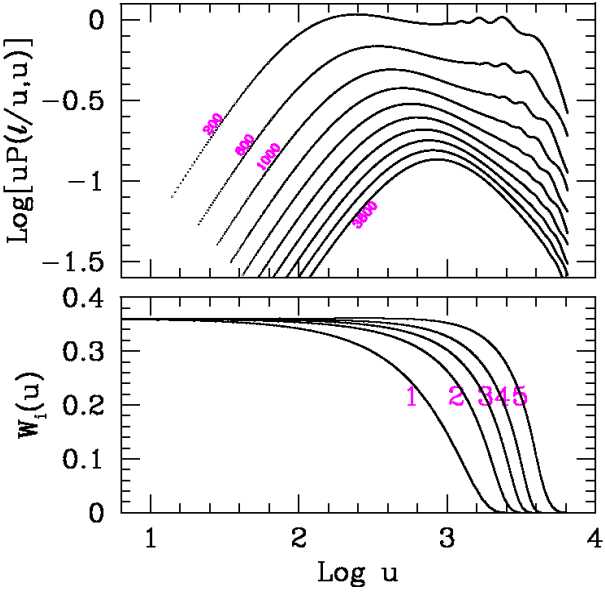}
\end{center}
\caption{Upper panel: Dependence on $u$ of the $P(\ell/u,u)$ integrand
  factor, for $\ell = 200,~400,~1000,~.....~,3800$. The curves are
  obtained by plotting the 10,000 points used in Riemann integration.
  Lower panel: window functions $W_i(u)$ in the five--bin case ($u$ in
  Mpc). 
 }
\label{wiz}
\end{figure}
\begin{figure}
\begin{center}
\includegraphics[scale=0.45]{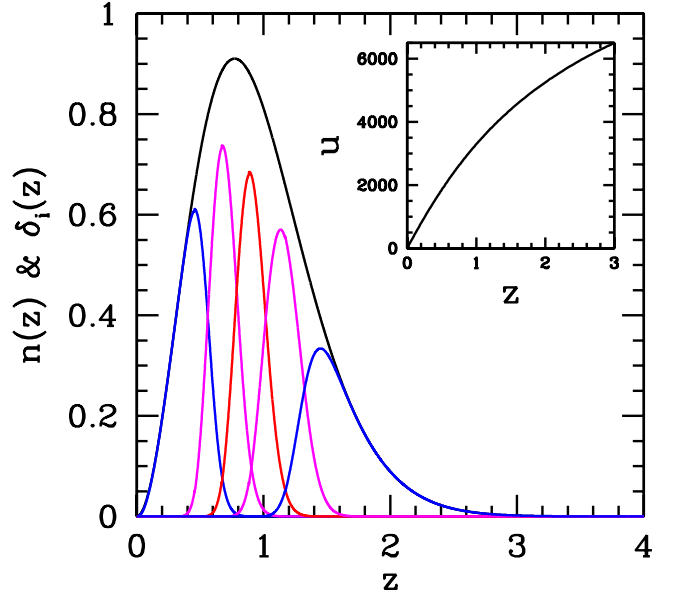}
\end{center}
\vskip -1.truecm
\caption{Assumed redshift distribution of galaxies and their expected
  partition among the five bins. In the inner panel we plot $u$ (in Mpc) 
versus ~$z$.
}
\label{EN5}
\end{figure}
\begin{figure}
\begin{center}
\includegraphics[height=9.cm,angle=0]{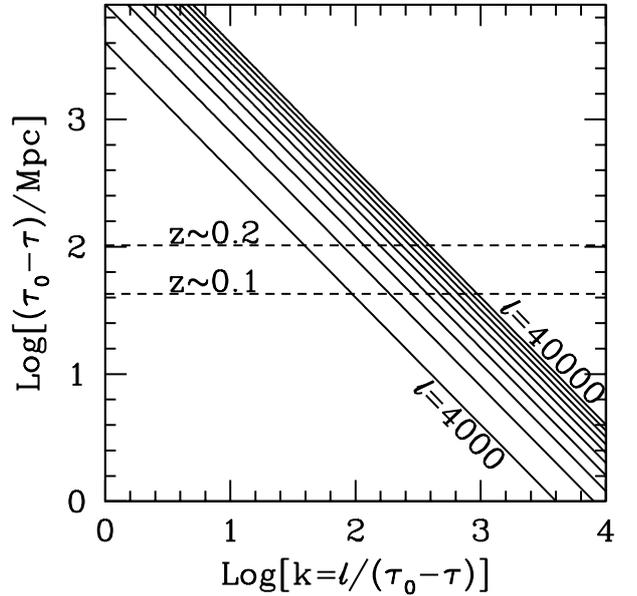}
\end{center}
\caption{Domains of integration of the matter power spectrum
  $P(k,\tau)$ on the $k$,$\tau_o-\tau$ plane, to compute the
  convergence spectrum. Each solid line corresponds to a value of $l$ ranging
  from $l=4000$ to $l=40000$ with increments of $\Delta l=4000$. The
  two horizontal dashed lines mark the value of the conformal time
  corresponding to redshifts $z=0.1$ and 0.2.}
\label{kl}
\vskip -.2truecm
\end{figure}
\begin{figure}
\begin{center}
\vskip -.3truecm
\includegraphics[height=9.cm,angle=0]{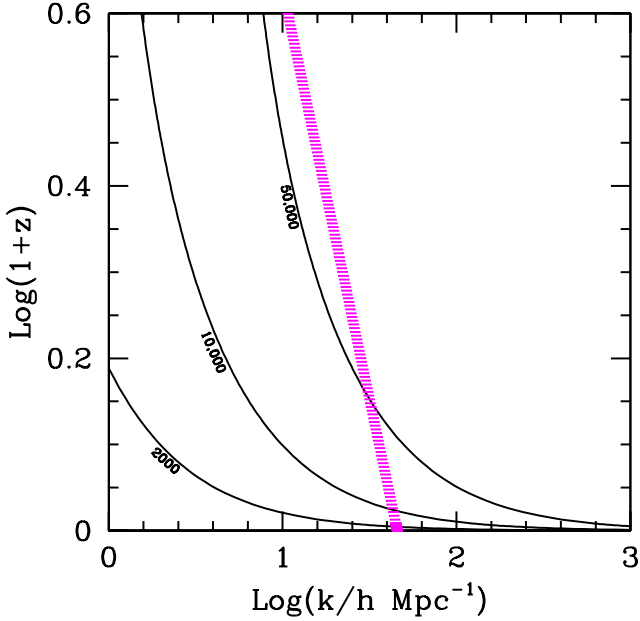}
\end{center}
\vskip -.7truecm
\caption{Numerical noise limit (magenta band) compared with the
  maximum $k$ needed to calculate the shear spectra at various $\ell$
  values (aside the curves). The $k$ values above which shot noise
  exceeds the physical spectrum is approximated here by the expression
  $\log(k/h\, {\rm Mpc}^{-1}) \sim 1.6 - 1.05 \, \log(1+z)$, in accordance
  with the results of the method described in Sec.~3.}  
\label{com}
\end{figure}
\begin{figure}
\begin{center}
\vskip -.6truecm
\includegraphics[height=9.cm,angle=0]{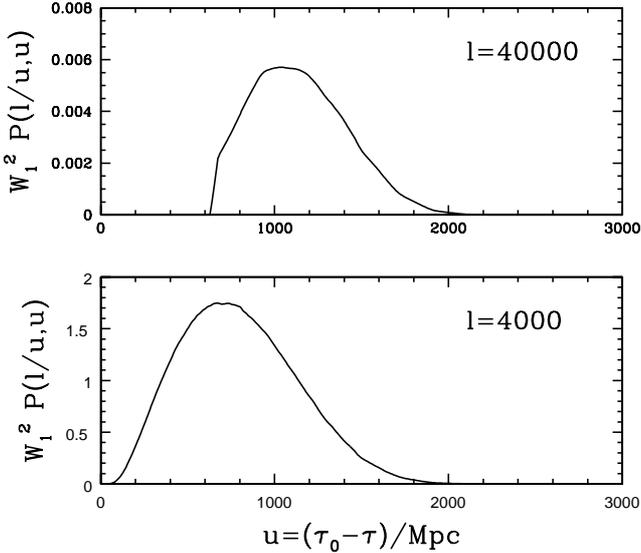}
\end{center}
\vskip -.4truecm
\caption{Dependence on $u=\tau_0-\tau$ of the integrand in
  eq.~(\ref{pijl}). The case $i=j=1$, which has the largest
  contribution at low redshift, is shown. The function is shown for
  $\ell = 4000$ (lower panel) and 40000 (upper panel).  We note the
  different ordinate scales and the low--$u$ cut in the latter case, where
   we estimate a reduction of $P_{11}(40000)$ by $\sim 4$
  per cent.}
\label{inte}
\vskip -.3truecm
\end{figure}
If we define
\begin{equation}
D_i(z) = n(z) \Pi_i(z)\,,
\label{diz}
\end{equation}
we can then introduce the distributions
\begin{equation}
\delta_{i}(z) = {D_{i}(z) \over \int_{0}^{\infty}D_{i}(z')dz'}\,,
\end{equation}
which indicate the actual redshift distribution of the galaxies
belonging to each bin (see Figure \ref{EN5}). From them,
we derive the filter functions
\begin{equation}
\label{WI}
W_i (z) = {3 \over 2} \Omega_m F_i(z) (1+z),
\end{equation}
where information about the galaxy redshift distribution is contained in
the quantity 
\begin{equation}
F_i (z) = \int_{\Delta z_i} dz'~\delta_i(z') u(z,z')/u(z')~.
\end{equation}
In the above expression, $u(z,z')$ is the (positive) conformal
time distance between $z$ and $z'$.  In Figure \ref{wiz}, we show the
$W_i$ profiles in the five--bin case; 
they are used in eq.~(\ref{pijl}), by replacing the dependence on $z$
with the dependence on the conformal time--distance~$u=\tau_0-\tau$
(see the inner box in the very Figure \ref{wiz}). Figure \ref{wiz}
highlights that in order to predict shear spectra we need fluctuation
spectra $P(k,z)$ up to $z \sim 2.5$, i.e. $u \simeq 5000$ Mpc.
In turn, this means that tomographic shear spectra for the assumed
redshift distribution will hardly provide sufficient information about the
evolution of the matter density power spectrum at $z >\sim 2~.$ The
redshift range $z=$0--2, however, is the one where one expects to be able 
to see the effect of the presence of DE.

The range of $k$ values over which we need to sample the matter power
spectrum is illustrated in Figure \ref{kl}. This figure shows how we can
derive shear spectra up to
$\ell \simeq 40000$ (and beyond), where we expect to have
underestimated them by $\mincir 4$ per cent, as we argue below.

The point is not to run the risk of using regions of the fluctuation
spectrum where numerical noise is significant. In Figure \ref{com}, we
then compare the numerical noise limit (magenta band) estimated from
power spectrum with $f=5$, with ~$k=\ell/u$, for $\ell = 2000,$ 10000
and 50000. The figure confirms that a danger might exist at small
$u$ values. However, even for $\ell=50,000$ the risks concern spectra
at $z <\sim 0.13$ and can be directly avoided by using spectra with
$f=4$, which, at such low $z$, are cut--off before the onset of
numerical noise. Rather than an overestimate, we then have an
underestimate of $P_{ij}(\ell)$ for large $\ell$ values. The 
underestimate is however small. For small $u$ values,
 the functions $W_i(u)W_j(u) P(l/u,u)$ provide negligible
contributions, as the window functions remains essentially constant, while
the spectrum decreases by several orders of magnitude, towards large
$k$'s, yielding a rapid decrease with $u$. This $u$--dependence is also
shown in the upper panel of Figure \ref{wiz} for a sample of small
$\ell$ values.

In Figure \ref{inte}, we then show the overall effect of the high--$k$
spectral cutoff for $i,j=1$, the spectrum most sensitive to low--$z$
contributions, by plotting the whole integrand $W_1(u)W_1(u) P(l/u,u)$
for $\ell = 4000$ and 40000~. For $\ell = 4000$, the function appears
quite regular, while for $\ell=40000$, we have a low--$u$ cut--off,
arising from the spectral cut--off. This cut leads to an underestimate of
$P_{11}(l)$, starting at $\ell \sim 25000$, reaching $\sim 4\, \%$ at
$\ell=40000$ (the case plotted) and attaining $\sim 10\, \%$ for
$\ell = 50000$.  Deficits are smaller at larger $i,j$ values.

\subsection{Tomographic shear spectra}
\begin{figure}
\begin{center}
\vskip -1.8truecm
\includegraphics[height=9.cm,angle=0]{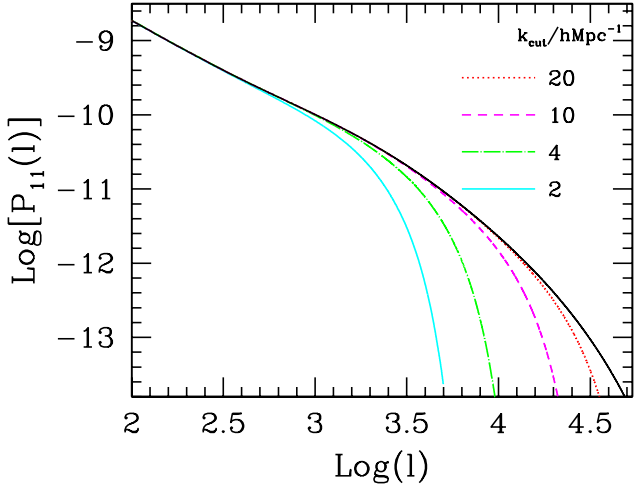}
\end{center}
\vskip -.7truecm
\caption{Effects on $P_{11}(\ell)$ of cutting the fluctuation
  spectra for the CSF simulations at different values of
  $k_{cut}$. Progressively lower curves correspond to increasing
  values of $k_{cut}$, as reported by the labels.
  }
\label{U1000}
\end{figure}
\begin{figure}
\begin{center}
\includegraphics[height=9.cm,angle=0]{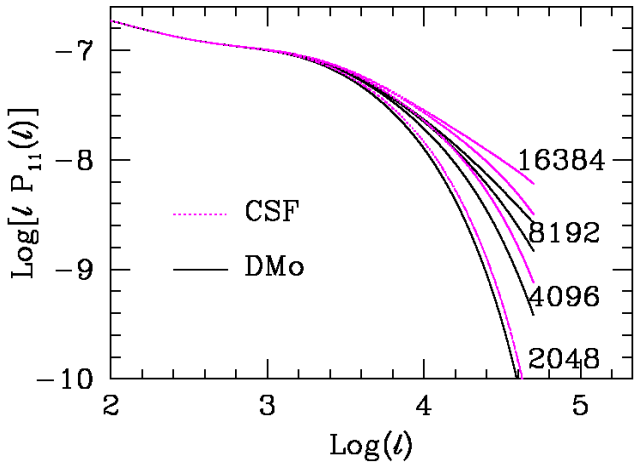}
\end{center}
\vskip -2.truecm
\caption{Large--$\ell$ cutoff of $ P_{11}(\ell)$, obtained by
  integrating spectra obtained with different grids. The effective
  number $n$ of grid points used is indicated in the frame. Note that
  the spectrum for the DMo simulation computed for $n=16384$ is lower
  than the spectrum of the CSF simulation computed for $n=8192$.}
\label{restest}
\end{figure}
Before using eq.~(\ref{pijl}) to evaluate the shear spectra up to
large $\ell$ value, we proceed with a few further tests.

In Figure \ref{U1000}, we show the effect of artificially cutting the
spectra at a redshift--independent value $k_{cut}$. When $k_{cut} =
50\, h$Mpc$^{-1}$, the logarithmic plot of $P_{11}$ does not appreciably
differ from the one computed using full spectra.
This plot can be compared with Figure \ref{restest}, which shows the
effects of different resolutions in the fluctuation spectra. This confirms
that lower resolution grids, yielding spectra that decline systematically
well before $k \sim 40\, h$Mpc$^{-1}$, are inadequate to reach large
$\ell$ values.

\begin{figure}
\begin{center}
\vskip .3truecm
\includegraphics[height=9.cm,angle=0]{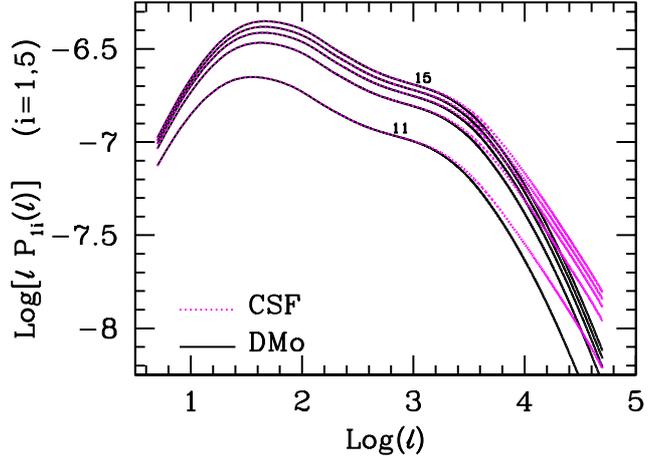}
\end{center}
\vskip -2.4truecm
\caption{Tomographic shear spectra in the five--bin case. In the
upper (lower) panel, we plot $P_{ii}(l)$ ($P_{1i}(l)$) with $i=1,...,5$.
for the hydrodynamical CSF and DMo N--body simulations.
}
\label{ii1i}
\end{figure}

\begin{figure}
\begin{center}
\includegraphics[height=7.cm,angle=0]{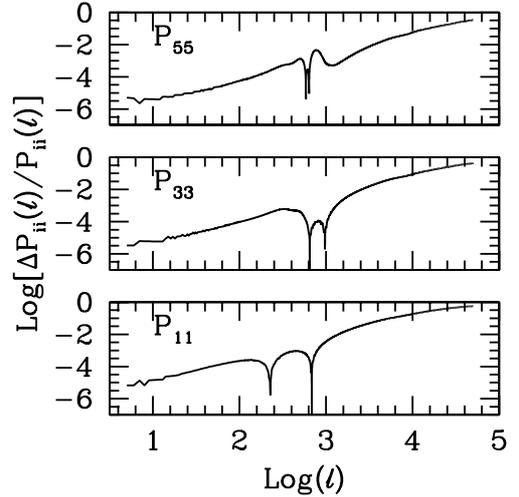}
\end{center}
\caption{Relative difference in the shear spectra for the CSF and the DMo
  simulations, for the three different values of $i,j$ reported in the
  labels. 
The sign of the shift changes in two points. Between them,
the logarithm of the modulus is plotted.}
\label{Dp}
\end{figure}

In Figure \ref{ii1i}, we then show the shear spectra $P_{ii}(\ell)$ and
$P_{1i}(\ell)$ for $i=1,...,5$, for both the hydrodynamical CSF simulation
and for DMo $N$--body simulation.
The extra power at large $k$'s in the spectrum of the CSF simulation
is clearly visible at large $\ell$ ($>1000$). The relative difference
in the power spectra between these two simulations can be more clearly
appreciated in Figure \ref{Dp}. This figure exhibits an interesting
feature: at $\ell \sim 10^2$--$10^3$, there is a short interval where
the shear spectra of the CSF simulation are slightly below the
corresponding ones of the DMo simulation. This interval is more
evident in $P_{ij}(\ell)$ for low $i,~j$, and is a consequence of the
power inversion in fluctuation spectra (see Figure 6). We point out however
that the difference between shear spectra is dominated by the large
shift from $P(k)$ at $k > \sim 2$--4~. When integrating over $u$,
such shifts affect all $\ell$. Hence, the detailed features of
Figure \ref{magnify} are diminished by the large--$k$ shift. When
comparing GH and DMo, these features appear more clearly (see below).

\begin{figure}
\begin{center}
\vskip -0.5truecm
\includegraphics[height=8.5cm,angle=0]{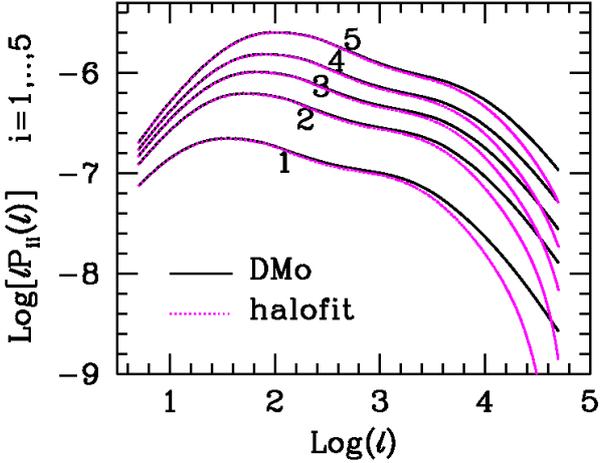}
\end{center}
\vskip -0.7truecm
\caption{N--body shear spectra compared with shear spectra obtained
  using HF, for the same model.  }
\label{camb}
\end{figure}

On the basis of the results shown in Figure 3, one may wonder to what accuracy
{\it HF} can reproduce simulation results for the shear spectra. 
This is shown in Figure \ref{camb}, which highlights a
major difference, that is sizable even at $\ell = 1000$.  This confirms
that using HF to deal with non--linear shear predictions can 
lead to substantial inaccuracies. 

\begin{figure}
\begin{center}
\vskip -.5truecm
\includegraphics[height=9.5cm,angle=0]{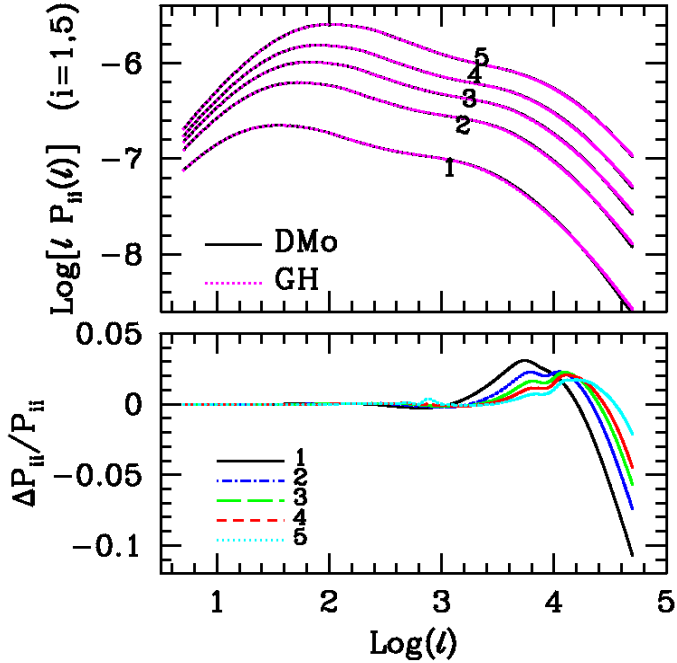}
\end{center}
\caption{Shear spectra from GH and DMo simulations nearly overlap. In
  the lower frame, we then plot the shift $(P_{ii(DMo)}-P_{ii(GH)})/
  P_{ii(GH)}$, which becomes significant only above $\ell \simeq
  1000$, as expected, and but remains within $10\, \%$, being
  smaller for larger $i$ values.  }
\label{redd}
\end{figure}

We finally test the effects of considering the effect of non--radiative
hydrodynamics on the shear spectra. In Figure \ref{redd},
we compare $P_{ii} (\ell)$ spectra for the GH hydrodynamics case with
$N$--body spectra, as done for CSF spectra in the top frame of Figure
\ref{ii1i}. We find that any difference between DMo and GH spectra is
small, never exceeding $10\, \%$ up to $\ell \simeq 50000$. An
interval where DMo spectra (marginally) are brighter than those of GH exists,
 between $\ell = 1000$ and $\sim 10000$. This inversion ceases at a higher
$\ell$ value for greater values of the indeces $i$. Altogether,
this plots reproduces, in terms of $\ell$, features that can be discerned 
when comparing $P(k)$ spectra, in Figure \ref{magnify}.

Before concluding this section, we comment on the accuracy of the
numerical integration yielding the shear spectra. Numerical inaccuracy
could cause a dangerous misunderstanding about the transfer of
fluctuation spectra features onto shear spectra. As a matter of fact,
all $P_{ij}(l)$ shown so far were obtained by performing the
integration in eq.~(2) using a simple Riemann algorithm, by summing
over $N_p \simeq 10^4$ integration points evenly distributed
between $u=0$ and 6500 Mpc ($z=0$ and 3).  For low $i,j$, quite a few
points are doomed to fall into a region where the window functions ensure 
that the integrand is negligible.  
For instance, Figure \ref{wiz} shows that $W_1^2 \sim
10^{-4}$ at $z \sim 0.5$ ($u \sim 1500~$Mpc); even in this worst case,
about $\sim 2400$ points are then significant. Convergence was however
numerically tested, by considering a variable $N_p$, and is already
satisfactory for $N_p = 2000~.$

\section{Conclusions}
We have presented results on the computation of density
fluctuation power spectra and tomographic cosmic--shear power spectra,
from a set of $N$--body and SPH hydrodynamical simulations of a
concordance $\Lambda$CDM model. Simulations were carried out in a box
with a comoving size of 410$\, h^{-1}$Mpc using $N=1024^3$ dark
matter (DM) particles and an equal number of gas particles in the
hydrodynamical simulations, with a Plummer--equivalent force
resolution of 7.5~$h^{-1}$kpc. This enabled us to compute spectra up
to $k \simeq 500\, h$Mpc$^{-1}$, whose reliability is only limited by
numerical noise effects that, even at $z=0$, are still evident for $k
> (50$--$70)\, h$Mpc$^{-1}$.

In addition to a DM--only N--body simulation, two hydrodynamical simulations
were carried out, the first one based on simple non--radiative
physics, and the second one including radiative cooling, star
formation, and the effect of SN feedback in the form of
galactic winds.  Since hydrodynamical simulations are based on twice
as many particles, their power spectra being characterized by
a relatively low level of high--$k$ white noise, with respect to the
DM-only simulation.

Taking advantage of the fairly large size of the simulation box, we
performed a direct connection between spectra from simulations and
linear--theory, in a $k$--range where the latter is expected to hold to
good accuracy.
A comparison with the predictions of {\it HF} 
confirms the unreliability of this algorithm in providing precise
predictions for non--linear power spectra. The deficit of power from
{\it HF} appears to be more and more severe at higher redshifts, likely
due to the delayed onset of non--linearity predicted by this approach.
When non--linearity is finally fully developed, discrepancies becomes
smaller, although are still quite significant.
This discrepancy could possibly arise from the lack of long--wave
contributions to the spectra used to build {\it HF}, that were
obtained from simulations in boxes whose side never exceeded $\sim
240\, h^{-1}$Mpc.
One cannot however exclude that the realization of this simulation
is notably above average.

This finding highlights the need to study the non--linearity onset by
using large enough boxes. In a number of recent papers studying the
effect of baryon physics on the non--linear power spectrum (e.g., Rudd
et al. 2008, Van Daalen et al. 2011), boxes with sides of mostly $\cal
O$$(60\, h^{-1}$Mpc) were used. Van Daalen et al. (2011), also used a
slightly larger box of $100\, h^{-1}$Mpc on a side.  In our opinion,
using boxes of such limited size can lead to non--numerically
converged estimates of the non--linear power spectrum.

In this paper, we have not debated the nature of DE, and referred to
purely $\Lambda$CDM cosmologies. Moreover, AGN feedback was not
  included,
at variance from
both Van Daalen et al. (2011) and Semboloni et al. (2011), 
who however made use of quite a smaller box.
Because of spectral discreteness,
to obtain shear spectra, they ought then to make recourse to approximated
spectral expressions, even before the spectral scale approaches the
box size, being there still far from a fully linear regime.
The whole spectral normalization may then become imprecise,
particulary when spectral expressions suitable to $\Lambda$CDM models
are tentatively extended to other cosmologies (see, e.g., Casarini el
al. 2011a). In contrast, our numerical spectra match a scale range
where linearity is substantially unaffected (see Figure
\ref{connect}). The box size is therefore a critical issue, as no
approximated expression is needed in our analysis.
 
As a matter of fact, following the onset of non--linearity without
introducing any source of bias, may become a key issue when 
  tomographic shear spectra of different DE models are compared. The
models often differ mostly in terms of the timing of the
non--linearity onset. When lacking the correct normalization between
spectra from simulations and linear theory, this timing is distorted
and model comparisons can be biased.

As in our study the effect of introducing baryons, our analysis confirms that
significant differences from $N$--body results appear at
large $k$ values when radiative physics is included. In this case, the
increase in small--scale power originates from the sinking of a
significant amount of cooled baryons in the central regions of DM
halos (e.g., Gnedin et al. 2004, Jing et al. 2006, Rudd et al. 2008,
Duffy et al. 2010, van Daalen et al. 2011).

As for the computation of a tomographic shear spectrum, 
we highlight that the large size of the box size used has allowed
us to obtain them from the density fluctuation spectra by directly
relating the linear spectrum to that measured from the simulations. In
most previous analyses, which are based on smaller simulation boxes,
{\it HF} expressions were used to cover a fairly wide
intermediate--range of $k$ values. 
The spectra $P_{ij}(\ell)$ could then be computed for $\ell$ values
reaching 50000. The simulations used here allowed us to achieve a
precision better than 1 per cent for $\ell <\sim 25000$, with an
underestimate of the spectrum amplitude of within 10 per cent at $\ell
\sim 50000$, in the worst case,  thanks also to pushing numerical
  noise interference up to $k > 40$--$50\, h\, $Mpc$^{-1}$.

Our results highlight the importance of accurately calibrating the subtle
effects of the propagation of signal from fluctuation spectra to shear
spectra, and the relevance of detailed large-scale simulations in
tracing the transition from linear to non--linear scales.

\vskip .3truecm

\noindent
ACKNOWLEDGMENTS. LC acknowledges the Brazilian research Institutions
FAPES and CNPq,  and the Observatory of Paris-Meudon
for their financial support. SAB acknowledges the
support of CIFS. We acknowledge partial support by the European
Commissions FP7 Marie Curie Initial Training Network CosmoComp
(PITN-GA-2009-238356), by the PRIN-INAF09 project “Towards an Italian
Network for Computational Cosmology”, by the PD51 INFN grant. We are
grateful to Volker Springel for making the non-public GADGET-3 code
available to us. Simulations were carried out at the CINECA
Supercomputing Centre in Bologna, with CPU time allocated through a
ISCRA proposal.

\vskip .3truecm
\vskip .3truecm


\vskip .2truecm

\noindent
{\bf Appendix A. Computation of power spectrum with folding}

\noindent
We consider a one--dimensional distribution of $N$ particles on a
segment of length $L$, i.e. with abscissas $0 < x_r < L~,$ given by
$$
\Delta (x) = \sum_{r=1}^N \delta(x-x_r)\,.
\eqno (A1)
$$
Its Fourier transform is
$$
P(k_n )
=
{1 \over \sqrt L} \int_0^L dx\,  \sum_{r=1}^N \delta(x-x_r) e^{2 \pi\, i\, nx/L} =
{1 \over \sqrt L}  \sum_{r=1}^N  e^{2 \pi\, i\, nx_r/L}
\eqno (A2)
$$ with $k_n = {2\pi \over L} \cdot n$. We then 
evaluate $P(k)$  using a simple FFT algorithm, obtaining the spectrum up
to $\bar n = 2^\mu$ ($\mu$ positive integer).
For instance, we consider $\mu = 9$, 
yielding
$\bar n = 512$ and therefore allowing a formal resolution down to the
scale $\lambda = L/512= L/2^9~.$

From $\Delta(x)$, we can obtain another distribution
$$
\Delta_2 (x)= \sum_{r=1}^N \delta(x-x'_r) = 
\sum_{x_r<L/2} \delta(x-x_r) +\sum_{x_r>L/2} \delta[x-(x_r-L/2)],
\eqno (A3)
$$ 
whose particles are all set on points $x'_r < L/2$, either because
they are originally so, or because their abscissa has been lowered by
$L/2~$.

Its Fourier transform is given by
$$
P(k_m )
=
{1 \over \sqrt {L/2}}  \left[\sum_{x_r<L/2}  e^{2 \pi\, i\, 2mx_r/L}
+
 \sum_{x_r>L/2}  e^{2 \pi\, i\, 2m(x_r-L/2)/L} \right]
\eqno (A4)
$$
with $k_m =  {2\pi \over L/2} \cdot m~$. The second term in the r.h.s.
however also reads
$$
 \sum_{x_r>L/2}  e^{2 \pi\, i\, 2m[x_r-L/2]/L}
=
 \sum_{x_r>L/2}  e^{2 \pi\, i\, 2m\, x_r/L - 2 \pi\, i\, m}
=
\sum_{x_r>L/2}  e^{2 \pi\, i\, 2m\, x_r/L },
$$
as $\exp [2 \pi\, i\, m] = 1$ for any integer $m$.

Accordingly, the spectra of $\Delta$ and $\Delta_2$ coincide, for $n=2m$,
apart of a factor $\sqrt{2}$. We again choose $\bar m = 2^\mu$ with $\mu=9$. 
The formal resolution scale is then $(L/2)/512 = L/2^{10}$. 
Accordingly, we double the resolution, although we evaluate the 
spectral harmonics only for even $n$ values.

The generalizations of this procedure to $L/4$, $L/8$, etc., and then
three dimensions are straightforward.

\end{document}